\documentstyle[11pt,newpasp,epsf]{article}
\input epsf

\newcommand{\EQ}{\begin{equation}}
\newcommand{\EN}{\end{equation}}
\newcommand{\EQA}{\begin{eqnarray}}
\newcommand{\ENA}{\end{eqnarray}}

{}
{}
{}
{}
{}
{}
{}

\newcommand{\nnn}{\hat{\mbox{\boldmath $n$}} {}}

\newcommand{\uu}{{\bf{u}}}
\newcommand{\BB}{{\bf{B}}}
\newcommand{\JJ}{{\bf{J}}}

\newcommand{\AAA}{{\bf{A}}}

\newcommand{\EE}{{\bf{E}}}

\newcommand{\nab}{\mbox{\boldmath $\nabla$} {}}

\newcommand{\dd}{{\rm d} {}}

\newcommand{\yjgr}[3]{ #1, {JGR,} {#2}, #3}

\newcommand{\yapj}[3]{ #1, {ApJ,} {#2}, #3}
\newcommand{\yapjl}[3]{ #1, {ApJ,} {#2}, #3}

\newcommand{\yan}[3]{ #1, {AN,} {#2}, #3}

\newcommand{\yana}[3]{ #1, {A\&A,} {#2}, #3}
\newcommand{\yanas}[3]{ #1, {A\&AS,} {#2}, #3}

\newcommand{\yjfm}[3]{ #1, {JFM,} {#2}, #3}

\newcommand{\yprl}[3]{ #1, {PRL,} {#2}, #3}

\newcommand{\ymn}[3]{ #1, {MNRAS,} {#2}, #3}

\newcommand{\ysph}[3]{ #1, {Solar Phys.,} {#2}, #3}
\newcommand{\ypr}[3]{ #1, {Phys. Rev.,} {#2}, #3}

\newcommand{\ybook}[3]{ #1, {#2} (#3)}
\newcommand{\yproc}[5]{ #1, in {#3}, ed. #4 (#5), #2}
\newcommand{\pproc}[4]{ #1, in {#2}, ed. #3 (#4), (in press)}

\begin{document}

\title{Helical surface structures}

\author{Axel Brandenburg}
\affil{Nordita, Blegdamsvej 17, DK-2100 Copenhagen \O, Denmark}

\author{Eric G.\ Blackman}
\affil{Department of Physics \& Astronomy, University of Rochester,
Rochester NY 14627}

\begin{abstract}

Over the past few years there has been growing interest in helical
magnetic field structures seen at the solar surface, in coronal mass
ejections, as well as in the solar wind. Although there is a great deal of
randomness in the data, on average the extended 
structures are mostly left-handed
on the northern hemisphere and right-handed on the southern. 
Surface field structures are also classified as dextral (= right bearing)  
and sinistral (= left bearing) occurring preferentially in the northern
and southern hemispheres respectively. 
Of particular interest here is a quantitative
measurement of the associated emergence rates of helical structures,
which translate to magnetic helicity fluxes. In this review, we 
give a brief survey of what has been found so far and what is expected based
on models. Particular emphasis is put on the scale dependence of the
associated fields and an attempt is made to estimate the helicity flux
of the mean field vs.\ fluctuating field.

\end{abstract}


\section{Introduction}

There is now good evidence for the helical nature of the solar
magnetic field. Early work by Seehafer (1990) suggested that fitting the
line of sight magnetograms of solar active regions to a linear (constant
alpha) force-free magnetic field yields systematically negative values
of alpha in the northern hemisphere and positive in the southern. 
Although the evidence for the hemispheric 
dependence was perhaps not completely convincing back then, 
subsequent work by different groups (Pevtsov, Canfield, \& Metcalf 1995; 
Rust \& Kumar 1996; Bao et al.\
1999; Pevtsov \& Latushko 2000) have confirmed the initial results.

The quantity being measured in these studies is usually the current
helicity, $\int\JJ\cdot\BB\,\dd V$, or the current helicity
density, $\JJ\cdot\BB$, where $\BB$ is the magnetic field strength,
$\JJ=\nab\times\BB/\mu_0$ is the current density, and $\mu_0$ is the
magnetic permeability. Of particular interest is actually the magnetic
helicity $\int\AAA\cdot\BB\,\dd V$, where $\AAA$ is the magnetic vector
potential with $\BB=\nab\times\AAA$.\footnote{In general, the boundary
of the volume is not a magnetic surface, i.e.\ $\BB\cdot\nnn\neq0$, and so
$\int\AAA\cdot\BB\,\dd V$ will not be gauge-invariant, i.e.\ the result
will be different if one redefines $\AAA\to\AAA+\nab\phi$, where $\phi$
is an arbitrarily chosen gauge potential. This is why one has instead
to use the relative magnetic helicity of Berger \& Field (1984).}

The magnetic helicity is of great theoretical interest because it satisfies a
conservation law: except for small resistive terms, 
its rate of change depends only on the gains and losses of magnetic
helicity through the boundaries. However, the magnetic helicity is a
volume integral which is probably hopeless to measure in practice,
because the field cannot be observed in the solar interior. What
is possible, however, is to measure surface-integrated magnetic
helicity fluxes of the form $\int(\EE\times\AAA)\cdot\dd V$, where
$\EE=\JJ/\sigma-\uu\times\BB$ is the electric field and $\sigma$ is the
electric conductivity.\footnote{Again, this quantity is gauge-dependent
and has to be substituted by an expression that is compatible with the
definition of the relative magnetic helicity of Berger \& Field (1984).}
Regardless of numerous complications, recent work has confirmed the basic
hemispheric dependence of the sign of both current helicity densities
and surface-integrated magnetic helicity fluxes (Berger \& Ruzmaikin 2000,
DeVore 2000, Chae 2000): negative in the north and positive in the south.

The connection between current helicity and dynamo theory was immediately
recognized. R\"adler \& Seehafer (1990) proposed that the observed
signs of the current helicity are characteristic of the small scale
field rather than the large scale field. From a dynamo point of view
this is not only plausible, but also desirable, as we shall explain
later. From an observational point of view this is far less obvious,
because the field on the scale of active regions and that associated
with coronal mass ejections (CMEs) is not generally understood as part
of the small scale field. Eclipse images of the sun map out quite clearly
the overall field line structure (e.g.\ Fig.~1 in Low 2001). From these
one sees that the field lines in helmet streamers above and around CMEs
merges naturally with the large scale of the sun. This is also seen in
soft X-ray images from Yohkoh (e.g.\ Fig.~7 in Low 2001).

The purpose of this paper is to point out that, on theoretical
grounds, one might expect a certain degree of simultaneous emergence of
small and large scale fields of opposite helicities within each hemisphere. 
(This concept was also discussed in Blackman \& Field 2000.)
In the following, we explain the reasoning behind such an  
expectation in light of recent work, and suggest 
possible observational signatures of the process.

\section{Simultaneous production of positive \& negative magnetic helicity}

Magnetic helicity is being produced by differential rotation and cyclonic
convection (the $\alpha$-effect). Both sources of magnetic helicity
have been discussed in the past (e.g.\ Berger \& Ruzmaikin 2000). Here
we focus on the effect of cyclonic convection ($\alpha$-effect). It is
well known that the $\alpha$-effect does not produce any net magnetic
helicity (so it obeys magnetic helicity conservation in the limit of large
magnetic Reynolds numbers). Instead, it produces simultaneously positive
and negative magnetic helicity associated with a spectral segregation
(Seehafer 1996, Ji 1999). The question then arises where
do each of these oppositely helical contributions of the magnetic field
go? There are three possibilities: reconnection across the equator,
resistive cancellation, and losses at the solar surface. The latter is
by far the most plausible one. What is not so clear is how exactly one
is supposed to picture the simultaneous loss of oppositely helical
magnetic fields. More importantly, why has there been no observational
evidence of this, neither quantitatively nor qualitatively?
Recently, however, D\'emoulin et al.\ (2002) reported that sufficiently
far away from the photospheric inversion line writhe helicity (resulting
from the relative rotation of opposite polarities) and twist helicity
(resulting from the intrinsic rotation of either polarity) can have
different signs. This may well be an indication of the anticipated
simultaneous loss of magnetic helicity of both signs. 

From a turbulence point of view, one expects that any kind of helical
stirring leads to the development of an inverse cascade (Pouquet, Frisch,
\& L\'eorat 1976). As is now well established from simulations, this can
be seen in power spectra of the magnetic energy: helical forcing at or
around some wavenumber $k_{\rm f}$ leads to a spectral bump at $k<k_{\rm
f}$ (larger wavelength) where the spectral magnetic helicity is opposite
to that at the forcing wavenumber. As time goes on, this bump travels
toward smaller $k$,  until it reaches the wavenumber corresponding to the 
scale of the system.

The inverse cascade mechanism and the $\alpha$-effect are similar
(but see Brandenburg 2002 for pointing out differences), and
they are widely considered to be the most plausible mechanism 
for explaining the solar magnetic field. In addition to the helicity
effect (inverse cascade or $\alpha$-effect), there is also shear (or
differential rotation) which amplifies the toroidal
magnetic field, regardless of magnetic helicity. A rough measure of the
relative importance of shear and helical turbulence can be obtained by
considering the ratio of toroidal to poloidal magnetic field. For the sun
this ratio is between 10 and 100. For poloidal magnetic field generation,
shear does not contribute.

Shear tends to produce large scale magnetic fields that oscillate on a
time scale long compared with the turnover time of the turbulence. This
result goes back to Parker (1955), and is well understood in the
framework of mean-field dynamo theory (Moffatt 1978, Parker 1979,
Krause \& R\"adler 1980, Zeldovich, Ruzmaikin, \& Sokoloff 1983),
and also confirmed using direct simulations of helical turbulence with
sinusoidal shear (Brandenburg, Bigazzi, \& Subramanian 2001). In the
framework of this model, the long term cycles are to be identified with
the 22-year magnetic cycle of the sun. The magnetic field takes the form
of traveling waves that migrate in the direction perpendicular to the
shear. This migration may be identified with the migration
of sunspot belts toward the equator, though 
under certain circumstances the direction of the
field migration can be overturned by meridional circulation (Choudhuri,
Sch\"ussler, \& Dikpati 1995, Durney 1995, K\"uker, R\"udiger, \&
Schultz 2001).

The outer boundaries of the sun do allow magnetic field to escape,
but it is not clear just how much magnetic flux really does
escape. In simulations of forced hydromagnetic turbulence with open
boundaries (pseudo-vacuum boundary conditions) magnetic field is found
to escape both on small and large scales, and these two contributions do
indeed have opposite signs of magnetic helicity, but the contribution
from small scales is found to be weak compared with that from larger
scales. It is not entirely clear yet whether the boundary condition
is realistic enough and whether the comparatively weak losses of small
scale field are representative of the real solar magnetic field.

Before we discuss why small scale losses of helical magnetic
fields are important (and even advantageous) for $\alpha$-effect dynamos,
we illustrate first how to picture such simultaneous losses of oppositely
helical magnetic fields, and what the observable signatures of this
process would be.

\section{Simultaneous losses of oppositely helical magnetic fields}

\begin{figure}[t]\epsfbox{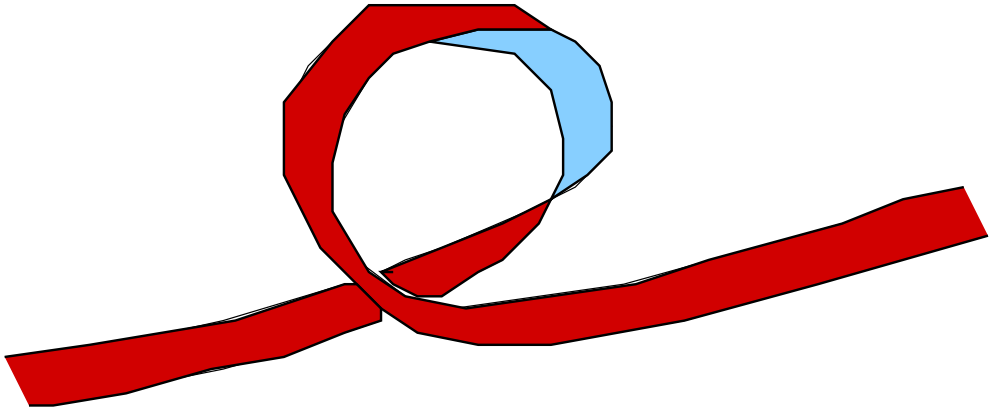}\caption[]{
Tilting of the rising tube due to the Coriolis force.
Note that the tilting of the rising loop causes also internal twist.
}\label{Fribbon2}\end{figure}

Given that magnetic helicity is conserved in the absence of boundary losses
and resistivity, any swirl-like motion must introduce simultaneously oppositely
helical magnetic fields when starting with an initially non-helical magnetic
field (Longcope \& Klapper 1997).
The prime example is of course the formation of an $\Omega$-shaped
flux loop due to magnetic or thermal buoyancy, and the simultaneous tilting
due to the Coriolis force. This is sketched in Fig.~\ref{Fribbon2}.

\epsfxsize=13.2cm

The tilting of the tube does clearly introduce current helicity,
$\JJ\cdot\BB$, where $\JJ$ is the current density associated with
the magnetic loop. The relation between this and the magnetic
helicity is not very direct. The resistive driving of the current
helicity is proportional to the current helicity,
\begin{equation}
{{\rm d}\over{\rm d}t}\int\AAA\cdot\BB\,{\rm d}V=
-2\eta\mu_0\int\JJ\cdot\BB\,{\rm d}V-\mbox{surface terms},
\end{equation}
but apart from this, the only direct relation is between the
spectra of magnetic and current helicities, $H(k)$ and $C(k)$,
respectively.\footnote{Spectra are straightforward to define
when the boundaries are periodic, so we restrict ourselves
only to this case here.} The spectra are normalized such that
\begin{equation}
\langle\AAA\cdot\BB\rangle=\int_0^\infty H(k){\rm d}k,
\end{equation}
where angular brackets denote volume averages, and
\begin{equation}
\langle\JJ\cdot\BB\rangle=\int_0^\infty C(k){\rm d}k;
\end{equation}
and the two are related to each other simply by
\begin{equation}
\mu_0 C(k)=k^2H(k).
\end{equation}
Since $H(k)$ and $C(k)$ can be of either sign, $\int H(k){\rm d}k$
can be of either sign for the same sign of  $\int C(k){\rm d}k$ 
for example. A useful tool is however the two-scale analysis, i.e.\ we define
$H_{\rm m}$ and $H_{\rm f}$ (and likewise $C_{\rm m}$ and $C_{\rm f}$)
as the contributions from mean and fluctuating field, corresponding
to the wavenumbers of the mean and fluctuating fields. Thus,
$H=H_{\rm m}+H_{\rm f}$ and $C=C_{\rm m}+C_{\rm f}$ with
\begin{equation}
\mu_0 C_{\rm m}=k_{\rm m}^2H_{\rm m},\quad
\mu_0 C_{\rm f}=k_{\rm f}^2H_{\rm f}.
\end{equation}
This immediately raises the question of whether the current helicity
generated by the rising flux tube is dominated by $k_{\rm m}$ or
by $k_{\rm f}$. In a sense the loop is of small scale by comparison
with the uniform field. On the other hand, when applied to the
regeneration of poloidal field from toroidal field, the newly
replenished poloidal magnetic field may well directly contribute
to the large scale magnetic field.

\begin{figure}[t]\epsfbox{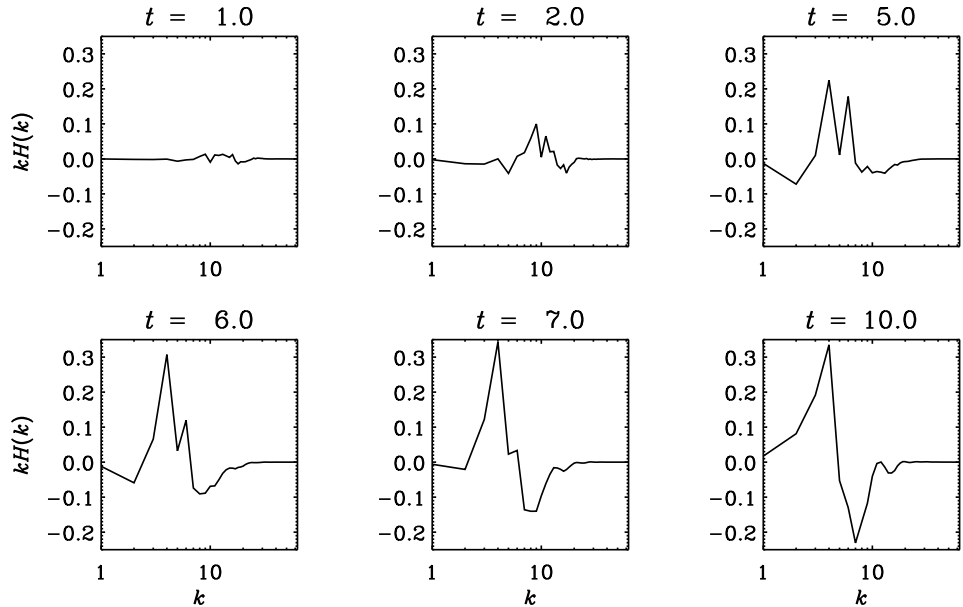}\caption[]{
Magnetic helicity spectra (scaled by wavenumber $k$ to give magnetic
helicity per logarithmic interval) taken over the entire computational
domain. The spectrum is dominated by a positive component at
large scales ($k=1-5$) and a negative component at small scales ($k>5$).
}\label{Fpspec}\end{figure}

We consider now the result of a simulation of a buoyant magnetic flux
tube. Similar calculations have been carried out many times in the past
(e.g.\ Abbett, Fisher, \& Fan 2000),
but here we are interested in the magnetic helicity spectrum which
does not seem to have attracted much attention so far. We start with a
horizontal flux tube in the azimuthal ($y$-) direction with vanishing
net flux (so there is a weak oppositely oriented field outside the tube)
and a $y$-dependent sinusoidal modulation of the entropy along the tube.
This destabilizes the tube such that it rises in one portion of the
tube.
Although the box is not periodic in the vertical direction, the boundary
conditions are still sufficiently far way so that we use Fourier transformation
to obtain power spectra of the magnetic helicity; see Fig.~\ref{Fpspec}.
Note that after some time ($t=6$ free-fall times) the spectrum
begins to show mostly positive magnetic
helicity (as expected), together with a gradually increasing higher
wavenumber component with the spectral helicity density is
negative. The latter is the anticipated contribution from small scales
resulting from the twist of the tube.

\epsfxsize=13.2cm

\begin{figure}[t]\epsfbox{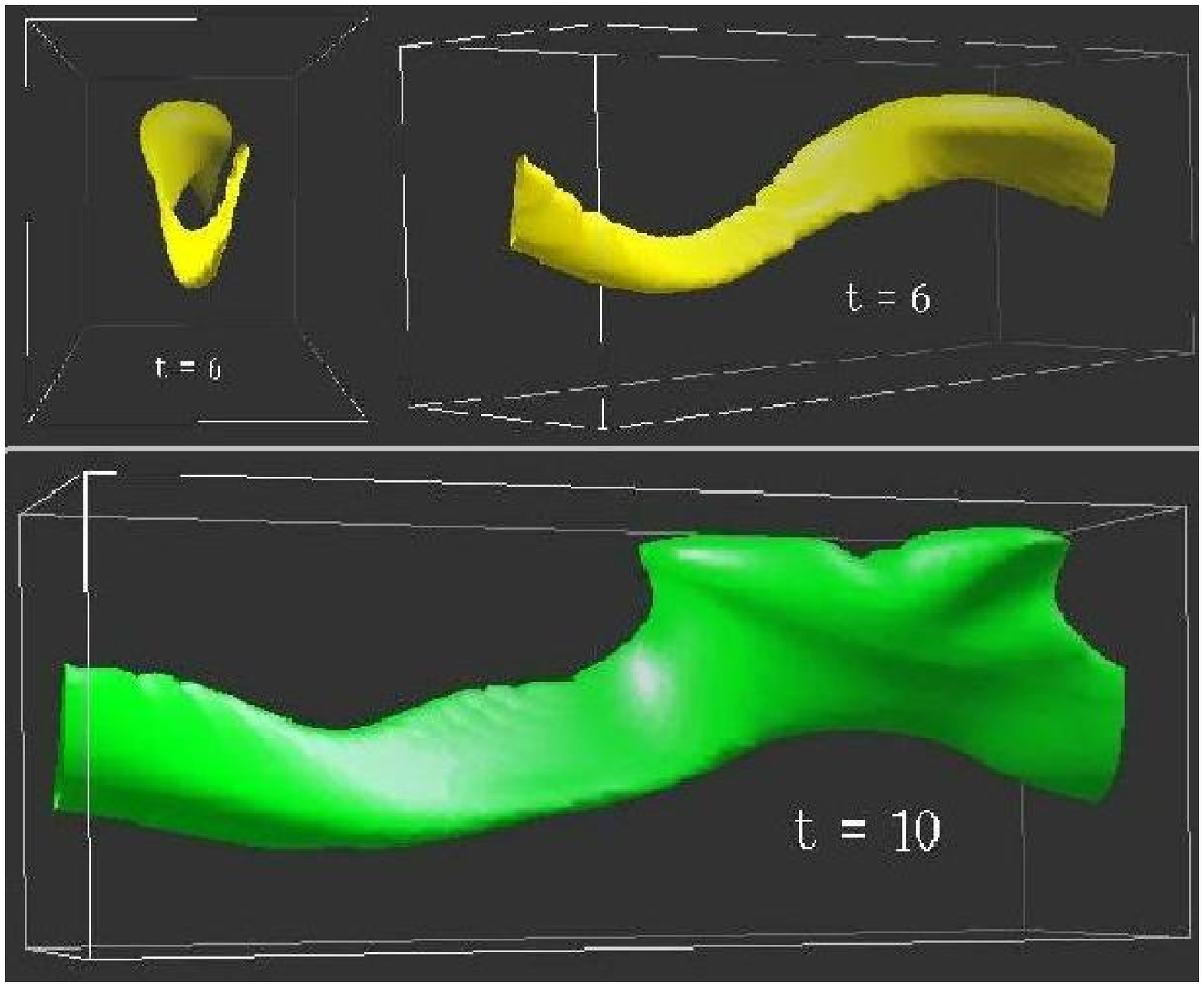}\caption[]{
Three-dimensional visualization of a rising flux tube in
the presence of rotation. The stratification is adiabatic such
that temperature, pressure, and density all vanish at a height
that is about 30\% above the vertical extent shown. (The actual
computational domain was actually larger in the $x$ and $z$ directions.)
}\label{Fall}\end{figure}

Instead of visualizing the magnetic field strength, which can be strongly
affected by local stretching, we visualize the rising flux tube using
a passive scalar field that was initially concentrated along the flux
tube. This is shown in Fig.~\ref{Fall}.

\epsfxsize=13.2cm

In future simulations we plan to follow the emergence of
the flux tube into the outer low plasma-beta exterior. We
expect that the losses of magnetic helicity have a scale
dependence that follows roughly that in the exterior.
In the following subsection we discuss the consequences of surface
losses of helical magnetic fields at small and large scales.

\section{Phenomenology of small and large scale field losses}

A relatively useful concept is based on the evolution equations for
small and large scale fields under the assumption that the fields are
maximally helical and have opposite signs of magnetic helicity at small
and large scales. The details can be found in Brandenburg, Dobler, \&
Subramanian (2002, Sect.~4.2). The strength of this approach is that
it is quite independent of mean-field theory.

Losses of large-scale field have been modeled using diffusion terms.
The phenomenological evolution equation are written in terms of the
magnetic energies and large and small scales, $M_{\rm m}$ and
$M_{\rm f}$, respectively, where we assume
$M_{\rm m}=\pm\mu_0 C_{\rm m}/k_{\rm m}$ and
$M_{\rm f}=\mp\mu_0 C_{\rm f}/k_{\rm f}$ for fully helical fields
(upper/lower signs apply to northern/southern hemispheres). The
phenomenological evolution equation take then the form
\begin{equation}
{{\rm d}M_{\rm m}\over{\rm d}t}=
-2\eta_{\rm m}k_{\rm m}M_{\rm m}
+2\eta_{\rm f}k_{\rm f}M_{\rm f},
\label{evolv_pheno}
\end{equation}
where $\eta_{\rm m}$ and $\eta_{\rm f}$ are effective magnetic
diffusivities that are expected to be anywhere between the molecular
magnetic diffusivity, $\eta$, and the turbulent magnetic diffusivity,
$\eta_{\rm t}$. The opposite signs with which $M_{\rm m}$ and
$M_{\rm f}$ enter reflect the fact that large and small scales contribute
with opposite signs. The case $\eta_{\rm m}=\eta_{\rm f}=\eta$ was already
discussed by Brandenburg (2001) who assumed that after a certain time
$t_{\rm sat}$, the small scale magnetic field will have saturated, so
$M_{\rm f}\approx\mbox{const}$ after $t>t_{\rm sat}$. After that time,
Eq.~(\ref{evolv_pheno}) can be solved and yields the solution
\begin{equation}
M_{\rm m}={\eta_{\rm f}k_{\rm f}\over\eta_{\rm m}k_{\rm m}}
\left[1-e^{-2\eta_{\rm m}k_{\rm m}^2(t-t_{\rm sat})}\right],
\quad\mbox{for $t>t_{\rm sat}$}.
\label{solution_pheno}
\end{equation}
This equation shows three things:
\begin{itemize}
\item
The time scale on which the large scale magnetic energy evolves
depends only on $\eta_{\rm m}$, not on $\eta_{\rm f}$.
\item
The saturation amplitude diminishes as $\eta_{\rm m}$ is
increased, which compensates the accelerated growth just past
$t_{\rm sat}$ (Brandenburg \& Dobler 2001).
\item
The reduction of the saturation amplitude due to $\eta_{\rm m}$
can be offset by having $\eta_{\rm m}\approx\eta_{\rm f}$, i.e.\
by having losses of small and large scale fields that are about
equally important.
\end{itemize}

The overall conclusion that emerges from this is, (i) $\eta_{\rm m}>\eta$
in order that the large scale field can evolve on a time scale other
than the resistive one, and (ii) $\eta_{\rm m}\approx\eta_{\rm f}$ in
order that the saturation amplitude is not catastrophically diminished.
These requirements are perfectly reasonable, but so far they have not been
borne out by simulations. Brandenburg \& Dobler (2001) found that most
of the losses of magnetic helicity occur on large scale. This is at first
glance very surprising, but on the other hand the magnetic helicity is a
quantity that is strongly dominated by the large scales. However,
certain phenomena such as CMEs and other perhaps less violent surface
events are not presently included in the simulations. As a proof of
concept, however, it has been possible to show that the artificial
removal of small scale magnetic fields (via Fourier filtering after a
certain number of time steps) can indeed lead to significant increase
of the saturation amplitude. 

The role of boundaries becomes particularly evident when considering
the fact that for closed or periodic boundaries the net flux through a
surface bounded by such boundaries cannot change. Indeed, the large scale
fields considered in Brandenburg (2001) also satisfy this property, so
the mean field is not simply the field averaged over the entire box, but
just horizontal averages.

\epsfxsize=13.2cm\begin{figure}[t]\epsfbox{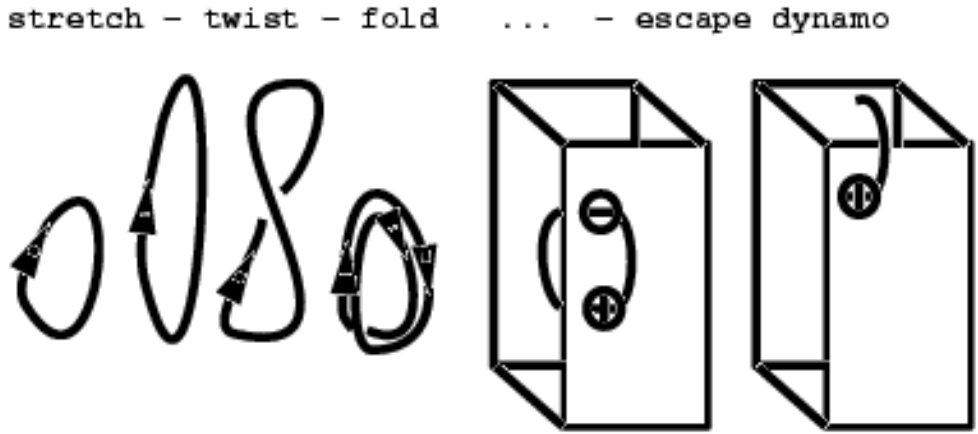}\caption[]{
Stretch-twist-fold (STF) dynamo with subsequent flux loss through
the upper boundary, leading to the production of net flux through the box.
(Adapted from Brandenburg 1998)
}\label{F_STFpicture}\end{figure}

The standard picture of a generic dynamo is the
stretch-twist-fold (STF) dynamo, which is depicted in Fig.~\ref{F_STFpicture}.
The flux through one half of the loop has doubled after one
STF iteration which, after gluing the two overlying loops
together, has lead to a configuration that is
topologically equivalent to the initial one. There is one
slight subtlety however: (a) after having twisted and folded
the two parts of the loop together we have simultaneously
introduced internal twist into the tube, very much like the
internal twist seen in Fig.~\ref{Fribbon2}. Again, this happened
only because magnetic helicity is such a well conserved
quantity, while still small and large scale magnetic helicity
have been introduced simultaneously.

As this twisted configuration goes through the boundary, magnetic
flux is lost partially, leaving a finite net magnetic flux
through the cross-section of the entire interior domain
(see the second part of Fig.~\ref{F_STFpicture}). At the
same time, though, no net magnetic helicity is lost
because the loop  simultaneously contains two canceling
contributions.  That such a loop has zero net helicity 
may be difficult to observe in practice  because the twist along the tube,
which corresponds to the small scale contribution, may be unresolvable 
if the overall structure is too small.
Another perhaps more plausible proposition is that the magnetic helicity
observed so far does already come from the small scales, and that it
is the large scale contribution that is not yet observed.

\section{Conclusions}

In this paper we have emphasized the importance of trying to detect
simultaneously large and small scale contributions to the losses of
helical magnetic fields at the solar surface. The motivation comes mostly
from isotropic turbulence simulations (similar high resolution
simulations of more realistic settings do not seem to be available yet),
but the basic reasoning is sufficiently general to warrant tentative
application to the sun. If our picture is correct, it would predict the
existence of an as yet unidentified helical component of the magnetic
field (with positive magnetic helicity in the northern hemisphere).
We expect that this unidentified component should be associated
with the large scale field rather than the small scale field. The
reason such a component is difficult to detect is related to the
fact that the large scale field is not seen directly. It only
manifests itself through the systematic orientation of
bipolar regions. However, such indirect indications have also been
used in the past to estimate the temporal--latitudinal behavior of
the large scale magnetic field from synoptic charts (Yoshimura 1976,
Stix 1976). This approach should be repeated with more complete
recent data to assess at least the sign of the large scale magnetic
helicity.

\acknowledgments
Use of supercomputer time on the PPARC supported machines at
Leicester and St Andrews, as well as the 512 node Beowulf cluster
in Odense is acknowledged.

\end{document}